\documentclass{iaus}
\usepackage{graphicx}

\title[Massive star evolution] %% give here short title %%
{Massive star evolution: from the early to the present day Universe}

\author[Georges Meynet et al.]   %% give here short author list %%
{Georges Meynet$^1$, Sylvia Ekstr\"om$^1$, Cyril Georgy$^1$ , Andr\'e Maeder$^1$, Raphael Hirschi$^2$}
\affiliation{$^1$Observatory of Geneva University, Switzerland \\ email: {\tt georges.meynet@obs.unige.ch}\\[\affilskip]
$^2$EPSAM, University of Keele, UK \\ email: {\tt r.hirschi@epsam.keele.ac.uk}}

\pubyear{2008}
\volume{252}  %% insert here IAU Symposium No.
\pagerange{119--126}
\jname{The art of stellar Modeling in the 21rst century}
\editors{L. Deng, K.L. Chan, C. Chiosi eds.}
\begin{document}

\maketitle

\begin{abstract}
Mass loss and axial rotation are playing key roles in shaping the evolution of massive stars. 
They affect the tracks in the HR diagram, the lifetimes, the surface
abundances, the hardness of the radiation field, the chemical yields, the presupernova status, the nature of the remnant,
the mechanical energy released in the interstellar medium,  etc...  In this paper, after recalling a few characteristics of mass loss and rotation, we review the effects of these two processes at different metallicities. Rotation probably has its most important effects at low metallicities, while mass loss and rotation deeply affect the evolution of massive stars at solar and higher than solar metallicities. 
\end{abstract}

\firstsection % if your document starts with a section,

\section{Mass loss due to radiative forces}

Radiation triggers mass loss through the line opacities in hot stars. 
It may also power strong mass loss through the continuum opacity when the star is near the Eddington limit. For cool stars, radiation pressure is exerted also on the dust.

For hot stars, typical values for the terminal wind velocity, $\upsilon_\infty$ is of the order
of 3 times the escape velocity, {\it i.e.} about 2000-3000 km/s,  mass loss rates are between 10$^{-8}$-10$^{-4}$ M$_\odot$ per year increasing with the luminosity and therefore the initial mass of the star (Vink et al. 2000; 2001). The comparison of mass loss rates for O-type stars obtained by different technics shows sometimes very important differences.
For instance, Fullerton et al. (2006) using UV line of P$^{+4}$ obtained mass loss rates reduced by  a factor ten or more with respect to mass loss determination from radio or H$\alpha$ determination. Bouret et al. (2005)
obtained qualitatively similar results to Fullerton et al. (2006) but with considerable lower reduction factor (about 3). Such reduction of the mass loss rates during the O-type star phase may have important consequences. Typically a 120 M$_\odot$ loses during its lifetime of 2.5 Myr about 50 M$_\odot$ whith mass loss rates of the order of 2 10$^{-5}$ M$_\odot$ per year. Dividing this mass loss rate by 10, would imply that in the same period, the star would lose only 5 M$_\odot$! Unless stars are strongly mixed (by e.g. fast rotation), or that all WR stars originate in binary systems, it would be difficult to understand how WR stars form with such low mass loss rates.
  
Stars with initial masses below about 30 M$_\odot$ at solar metallicity evolve to the red supergiant stage where mass loss is enhanced with respect to the mass loss rates in the blue part of the HR diagram (see for instance de Jager et al. 1988). In this evolutionary stage, determination of the mass loss is more difficult than in the blue part of the HR diagram due in part to the presence of dust and to
various instabilities active in red supergiant atmospheres (e.g. convection becomes supersonic and
turbulent pressure can no long be ignored). An illustration of the difficulty comes from the determination of red
supergiant mass loss rates by van Loon et al. (2005). Their study is based on the analysis of optical spectra of a sample of dust-enshrouded red giants in the LMC, complemented with spectroscopic and infrared photometric data from the literature. Comparison with galactic
AGB stars and red supergiants shows excellent agreement for dust-enshrouded objects, but not
for optically bright ones. This indicates that their recipe only applies to dust-enshrouded stars.
If applied to objects which are not dust enshrouded, their formula gives 
values which are overestimated by a factor 3-50! In this context the questions of which stars do
become dust-enshrouded, at which stage, for how long, become critical to make correct implementations of such mass loss recipes in models.

Stars with initial masses above about 30 M$_\odot$ at solar metallicity may evolve into a short Luminous Blue Variable (LBV) phase. 
LBV stars show during outbursts mass loss rates as high as 10$^{-4}$-10$^{-1}$ M$_\odot$ per year.
For instance $\eta$ Carinae ejected near the middle of the eighteenth century
between 12 and 20 M$_\odot$ in a period of 20 years, giving an average mass loss rate during this period of 0.5 M$_\odot$ per year.
Such a high mass loss cannot be only radiatively driven according to Owocki et al. (2004). These authors have shown that the maximum mass loss rate that radiation can drive is given by
$\dot M\sim 1.4 \times 10^{-4} L_6 {\rm M}_\odot {\rm yr}^{-1},$
with $L_6$ the luminosity expressed in unit of 10$^6$ L$_\odot$. 
This means that for $L_6=5$ (about the case of $\eta$Car) the maximum mass loss rate would be less than 10$^{-3}$ M$_\odot$ per year, well below the mass loss during the outbursts. These outbursts, which are more shell ejections than steady stellar winds, involve other processes in addition to the effects of the radiation pressure. Among the models proposed let us mention the geyser model by Maeder (1992b), or the reaching of the $\Omega\Gamma$-limit (Maeder \& Meynet 2000a).

After the LBV phase, massive stars evolve into the Wolf-Rayet phase, also characterized by strong mass loss rates. Many recent grids of stellar models use the recipe given by Nugis \& Lamers (2000) for the WR mass loss rates. These authors deduced the mass loss rates from radio emission power and accounted for the clumping effects.

%\begin{figure}
%\includegraphics[width=2.6in,height=2.6in,angle=0]{jm85z00S800.eps}
%\hfill
%\includegraphics[width=2.6in,height=2.6in,angle=0]{om85z00S800.eps}
%\caption{\textit{Left:} profile of the specific angular momentum $j_\mathrm{m}$ inside our 85 $M_\odot$ at the end of hydrostatic core Si-burning (continuous line). The dotted line is $j_{\rm K}=r_{\rm LSO}\,c$ \cite[ p. 428]{ST83}, where the radius of the last stable orbit, $r_{\rm LSO}$, is given by $r_{\rm ms}$ in formula (12.7.24) from \cite[p. 362]{ST83} for circular orbit in the Kerr metric. $j_{\rm K}$ is the minimum specific angular momentum necessary to form an accretion disc around a rotating black hole. $j_\mathrm{Schwarzschild}=\sqrt{12}Gm/c$ (long-dashed line) and $j_\mathrm{Kerr}^\mathrm{max}=Gm/c$ (short-dashed line) are the minimum specific angular momentum necessary for a non-rotating and a maximally-rotating black hole, respectively. \textit{Right:} profile of $\Omega_\mathrm{m}$ inside the same model.
%}\label{OME}
%\end{figure}

\subsection{Metallicity dependence of the stellar winds}

In addition to the intensity of the stellar winds for different evolutionary phases, one needs to know 
how the winds vary with the metallicity. This is a key effect to understand the different massive star populations observed in regions of different metallicities. This has also an important impact on the nature of the stellar remnant and on the chemical yields expected from stellar models at various metallicities .

Current wisdom considers that very metal-poor stars lose no or very small amounts
of mass through radiatively driven stellar winds. This comes from the fact that when the metallicity
is low, the number of absorbing lines is small and thus the coupling between the radiative forces and the matter is weak. Wind models impose a scaling relation of the kind
$
\dot M(Z)=\left({Z \over Z_\odot} \right)^\alpha\dot M(Z_\odot),
$
where $\dot M(Z)$ is the mass loss rate when the metallicitity is equal to $Z$ and $\dot M(Z_\odot)$
is the mass loss rate for the solar metallicity, $Z$ being the mass fraction of heavy elements.
In the metallicity range from 1/30 to 3.0 times solar,
the value of $\alpha$ is between 0.5 and 0.8 according to stellar wind models (Kudritzki et al. 1987; Leitherer et al. 1992; Vink et al. 2001). 
Such a scaling law implies for instance that 
a non-rotating 60 M$_\odot$ with $Z=0.02$
ends its stellar life with a final mass of 14.6 M$_\odot$, the same model with a metallicity of
$Z=10^{-5}$ ends its lifetime with a mass of 59.57 M$_\odot$ (cf. models of Meynet \& Maeder 2005 and Meynet et al. 2006 with $\alpha=0.5$).

During the red supergiant stage, at the moment there is no commonly accepted rule to account for a possible metallicity dependence of the winds. let us just mention here that according to
van Loon et al. (2005),  dust-enshrouded objects mass loss appears to be similar for objects in the LMC and the Galaxy. This may have very important consequences for our understanding of metal-poor red supergiant stars. For the LBV's, there is also no real knowledge on how mass loss can depend on metallicity. If the mechanism is mainly triggered by 
continuum opacity, we can expect that there is only a weak or may be no dependence on the metallicity.

Until very recently, it was considered that the WR mass loss rates did not depend on the initial metallicity {i.e.} that a WN stars in the SMC, LMC and
in the Galaxy would lose mass at the same rate provided they have the same luminosity and the same actual surface abundances. This view has been challenged by Vink \& de Koter (2005) who find that the winds of WN stars are mainly triggered by iron lines. They suggest a dependence of mass loss on $Z$ (initial value) similar to that of massive OB stars. According to these authors, the winds of WC stars depends also on the iron abundance, but in this case, the metallicity dependence is less steep than for OB stars.
Their results apply over a range of metallicities given by 10$^{-5} \le (Z/Z_\odot) \le 10$. 
Very interestingly, they find that once the metal abundance drops below $(Z/Z_\odot) \sim 10^{-3}$,
the mass loss of WC stars no longer declines. This is due to an increased importance of radiative driving by intermediate-mass elements, such as carbon. These results have profound consequences for the evolution of stars at low metallicity, affecting the predicted Wolf-Rayet populations
(Eldridge \& Vink 2006), the evolution of the progenitors of collapsars and long soft Gamma Ray Bursts (Yoon \& Langer 2005; Woosley \& Heger 2006ab; Meynet \& Maeder 2007).
  
\section{Rotation}

Rotation induces many processes in stellar interior (see the review by Maeder \& Meynet 2000a).
In particular, it drives instabilities which transport angular momentum and chemical species.
Assuming that
the star rapidly settles into a state of shellular rotation (constant angular velocity 
at the surface of isobars), the transport equations due to meridional currents and shear instabilities
can be consistently obtained (Zahn 1992). Since the work by J.-P.~Zahn, various improvements have been brought to the
formulas giving the velocity of the meridional currents (Maeder \& Zahn 1998), those of the various diffusive coefficients 
describing the effects of shear turbulence (Maeder 1997; Talon \& Zahn 1997; Maeder 2003; Mathis et al. 2004), as well as the effects of rotation on the mass loss (Owocki et al. 1996; Maeder 1999; Maeder \& Meynet 2000b). 

Let us recall a few basic results obtained from rotating stellar models:

1) Angular momentum is mainly transported by the meridional currents. 
During the Main-Sequence phase, the core contracts and spins up and the envelope expands and spins down. The meridional currents
impose some coupling between the two, slowing down the core and accelerating the outer layers.
In the outer layers, the velocity of these currents becomes smaller when the density gets higher.
As a consequence, the transport of angular momentum from inner to outer regions is less
efficient at low metallicity where stars are more compact and thus more dense in the outer layers.

2) The chemical species are mainly transported by shear turbulence (at least in absence of
magnetic fields; when magnetic fields are amplified by differential rotation as in the Tayler-Spruit
dynamo mechanism, Spruit 2002 , the main transport mechanism is meridional circulation, Maeder \& Meynet 2005). 
This process may produce changes of the surface abundances of rotating stars already during
the Main-Sequence phase (Heger \& Langer 2000; Meynet \& Maeder 2000).
The shear turbulence is stronger when the gradients of the angular velocity are stronger, {\it i.e.} at low metallicity where stars are more compact and where the meridional currents are slower. 

In addition to these internal transport processes, rotation also modifies the physical properties
of the stellar surface. Indeed the shape of the star is deformed by rotation (a fact which is now put in evidence 
observationally thanks to the interferometry, see Domiciano de Souza et al. 2003). Rotation implies a non-uniform brightness (also now
observed, see e.g. Domiciano de Souza et al. 2005).
The polar regions are brighter than the equatorial ones. This is a consequence of the hydrostatic
and radiative equilibrium (von Zeipel theorem 1924). The von Zeipel theorem has many very interesting consequences, among them let us mention the followings (see Maeder 1999; Maeder \& Meynet 2000b):
\begin{itemize}
\item Since the radiative flux and the effective gravity varies as a function of the colatitude, one can define a local Eddington factor as the ratio of the actual flux at that colatitude to the maximum flux allowed at that colatitude (the maximum flux is defined as the flux for which the radiative acceleration compensate for the effective gravity).
\item The critical velocity, defined as the value of the rotation velocity at the equator such that the centrifugal acceleration compensates for the net radial attracting force (which results from
the gravity in part counterbalanced by an outward radiative acceleration) is different when the
stellar luminosity is near or far from the Eddington luminosity.
\item The mass loss rates per unit surface varies as a function of the colatitude. Rotation induces wind anisotropies. Polar winds are expected in fast rotating hot stars.
\item The line driven mass loss rates are enhanced by rotation.
\end{itemize}

\subsection{Dependence on metallicity of effects induced by rotation}

\begin{figure}
\includegraphics[width=2.6in,height=2.6in,angle=0]{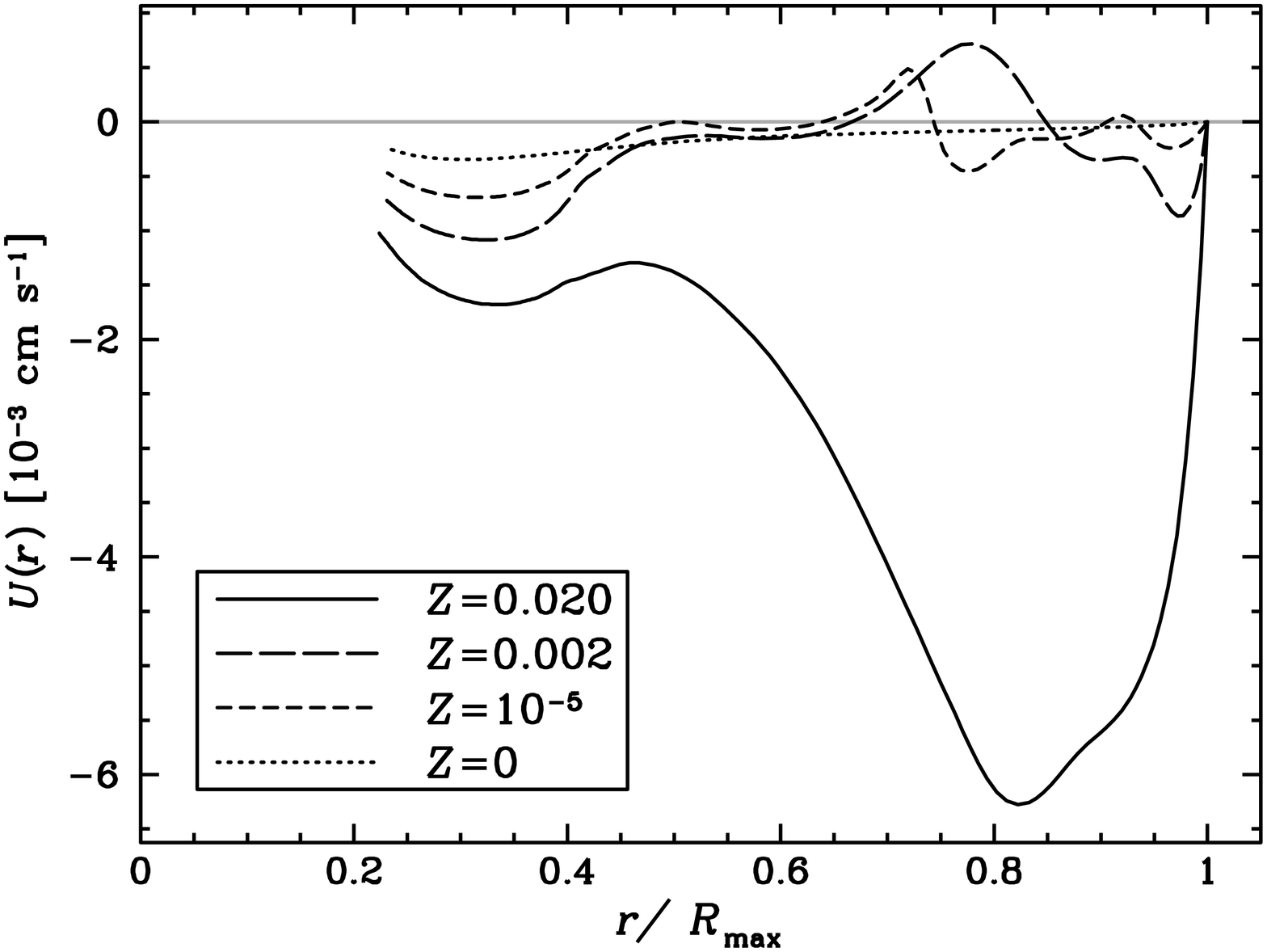}
\hfill
%\hspace{1cm}
\includegraphics[width=2.6in,height=2.6in,angle=0]{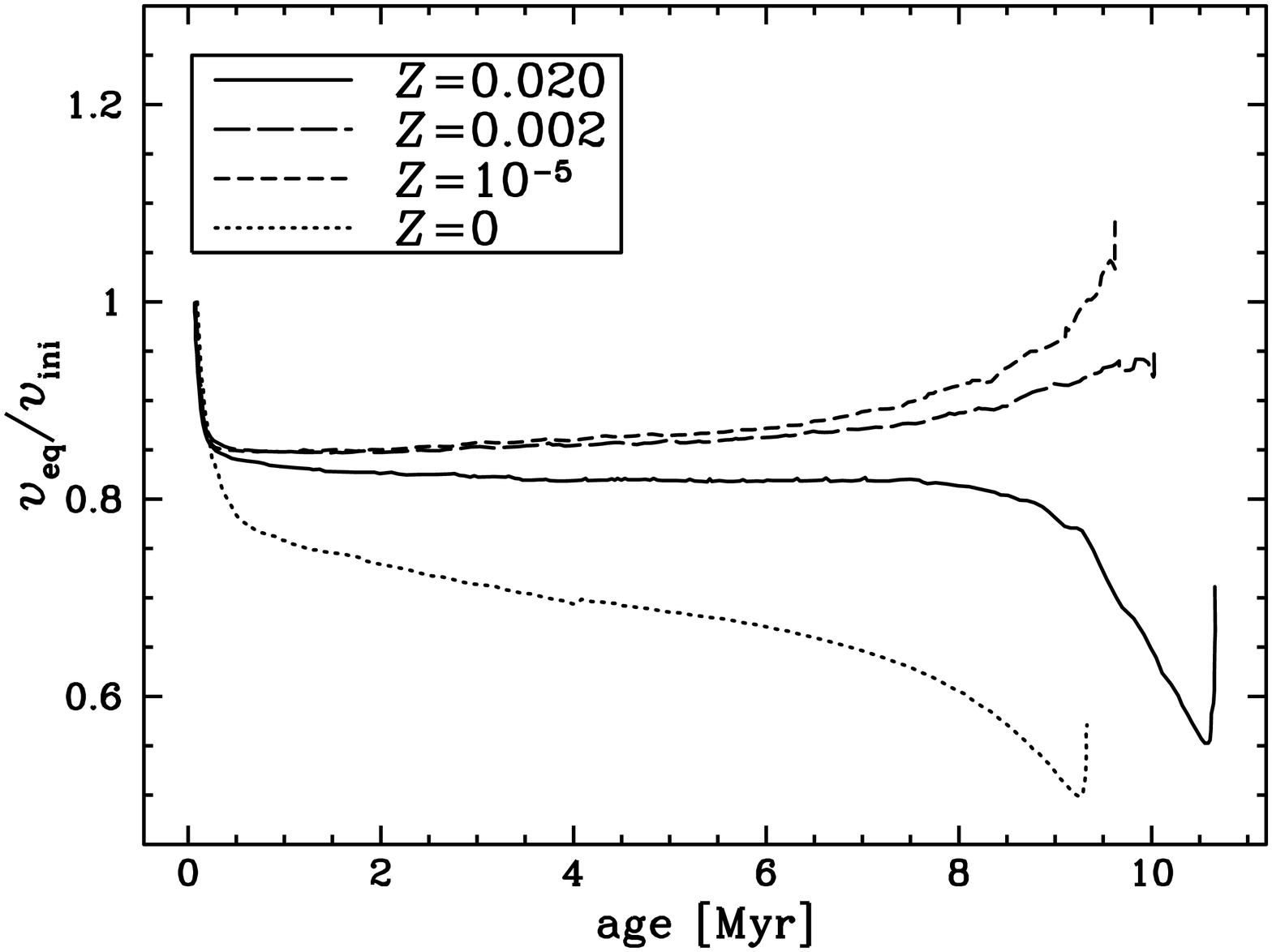}
\caption{Models of 20 M$_\odot$ at various metallicities, with $\Omega_\mathrm{ini}/\Omega_\mathrm{crit}= 0.5$. {\it Left:} internal profile of $U(r)$, where $u(r,\theta)$ the vertical component of the velocity of the meridional circulation is $u(r,\theta)= U(r)\ P_2\cos \theta$. The radius is normalized to the outer one. All the models are at the same evolutionary stage, when the central H mass fraction is about 0.40. {\it Right:} evolution of the equatorial velocity, normalized to the initial velocity. Figure taken from Ekstr\"om et al. (2008b).
}\label{furvevolz}
\end{figure}

The effects of rotation are not the same at different metallicities. These differences may have different causes:
\begin{enumerate}
\item The distribution of the rotational velocities on the ZAMS may depend on the metallicity.
Presently, observations seem to favor high rotation at low $Z$.
For instance, the observed number fraction of Be stars with respect to the total number of B stars is higher at low than at high metallicity (Maeder et al. 1999; Wisniewski \& Bjorkman 2006). 
Be stars are supposed to rotate near the critical velocity and are characterized from an observational point of view by the presence of an equatorial disk expanding outwards where emission lines (responsible for the ``e'' in Be name) are formed. Another argument supporting the greater number of fast rotators at low metallicity is the fact that the observed surface velocity of stars on the Main-Sequence do appear to be higher at low metallicities. For instance Hunter et al. 2008a obtain that SMC metallicity stars rotate on average faster than galactic ones (mainly field objects). No difference is found between galactic and LMC stars. Martayan et al. (2007a) find that, for B and Be stars, the lower the metallicity, the higher the rotational velocities (see the review by Meynet et al. 2008 for a more complete review of recent results from large massive star surveys).
Let us note that part of this effect may be attributed not to a difference in the initial velocity distribution but to the fact that at low metallicity mass loss rates are weaker (see above) and thus remove small amounts of angular momentum from the star.

Let us note 
that the mechanisms which evacuate the angular momentum from the protostellar cloud during the
stellar formation phase may have different efficiencies at low and high metallicity. For instance at high metallicity, the fraction of chemical species easily ionized is higher. This might imply a stronger magnetic coupling between the disk and the star (disk-locking effect) during the pre-MS phases and thus a more efficient spin-down. At low metallicity, it may also occur that the disk is more rapidly photoevaporate by the strong ionizing flux of young massive stars in the vicinity of the nascent star. Indeed, when the metallicity decreases, stars are bluer. This effect may make the disk-locking phase shorter and allow the nascent star to retain more angular momentum\footnote{In that context it is interesting to note that the rotational velocities of massive stars in clusters is higher than in the field supporting the view that the environment plays also a role in shaping the rotational velocity distribution.}. 

\item The evolution of the surface (equatorial) velocity results of a delicate interplay between the mass loss (which removes angular momentum at the surface) and the meridional circulation (which brings angular momentum from the core to the surface). 
In the left panel of Fig.~\ref{furvevolz}, we show the internal profile of the radial component of the meridional circulation in four models of 20 M$_{\odot}$ with $\Omega_\mathrm{ini}/\Omega_\mathrm{crit}= 0.50$ at various metallicities\footnote{These models are taken from Ekstr\"om et al. (2008a)} . Let us focus on the outer cell, which transports the angular momentum outward when $U(r)$ is negative. The amplitude of the meridional circulation is a factor 6 higher in the standard solar metallicity model ($Z=0.020$) compared to the $Z=0.002$ model. This factor amounts to 25 when we compare with the $Z=10^{-5}$ model, and reaches 100 with the $Z=0$ one. This illustrates the effect of the Gratton-\"Opik term ($\propto 1/\rho$) in the expression of the meridional circulation velocity $U(r)$: when the metallicity decreases, stars are more compact, so the density increases, and thus $U(r)$ decreases. In the right panel of Fig.~\ref{furvevolz}, we see the resulting evolution of the equatorial velocity. At standard metallicity, although the amplitude of the meridional circulation is large, the loss of angular momentum through the radiative winds has the strongest effect, and the equatorial velocity slows down. At low or very low $Z$, the meridional circulation is weak, but the mass loss is so diminished that the models are spinning up. In the case of $Z=0$ strictly, there is no mass loss to remove mass and angular momentum, but the meridional circulation is so weak that the evolution of $\Omega(r)$ is very close to local angular momentum conservation, $\Omega r^2 = \mathrm{constant}$: because of the natural inflation of the external radius, the surface of the model has to slow down.
\item A consequence of the above metallicity dependence is that the gradient of the angular velocity in a low metallicity stellar model is steeper than in a corresponding model at high metallicity.
Since the shear instability responsible for the mixing of the chemical mixing is stronger when the gradient of $\Omega$ is stronger, this implies that a given initial mass stars at low metallicity will undergo more mixing than a similar star at high metallicity.  
\end{enumerate}
In the following we explore the consequences of rotating stellar models with mass loss in different metallicity environments.

\section{Evolution of massive stars with mass loss and rotation at $Z=0$}

Recently Ekstr\"om et al. (2008b) computed a set of rotating Pop III massive star models.
These authors chose as initial equatorial velocity, a value  of 800 km s$^{-1}$ which corresponds to an angular momentum content approximately equivalent to that of massive star models at solar metallicity with rotation velocities of about 300 km $s^{-1}$. Let us recall that the value
of 300 km $s^{-1}$ on the ZAMS produce models with a time averaged velocity on the MS equal to 200 and 250 km s$^{-1}$, {\it i.e.} well in the range of observed values (see e.g. Huang \& Gies 2006ab). 

Comparing the evolution with and without rotation for Pop III stellar models, the following main differences can be noted:
\begin{itemize}
\item At $Z=0$ models rotate with an internal profile $\Omega(r)$ close to local angular momentum conservation, because of a very weak core-envelope coupling.
\item Rotational mixing drives a H-shell boost due to a sudden onset of CNO cycle in the shell, which leads to high $^{14}$N production. This production can be as much as $10^6$ times higher than the production of the non-rotating models. Generally, the rotating models produce much more metals than their non-rotating counterparts.
\item The mass loss is very low, even for the models that reach the critical velocity during the main sequence. It may however have an impact on the chemical enrichment of the Universe, because some of the stars are supposed to collapse directly into black holes, contributing to the enrichment only through their winds. While in that case non-rotating stars would not contribute at all, rotating stars may leave an imprint in their surrounding.
\item Due to the low mass loss and the weak coupling, the core retains a high angular momentum at the end of the evolution. The high rotation rate at death probably leads to a much stronger explosion than previously expected, changing the fate of the models.
\end{itemize}

None of our models meet the conditions for becoming WR stars. They end their evolution having kept their envelope, and thus seem to fail in becoming a GRB progenitor as defined in the collapsar model of Woosley (1993). However, we have seen that the rotating 85 $M_{\odot}$ is likely to undergo pulsational pair instability\footnote{This process has not been followed in the present models.} and thus lose some mass in this process. If the mass lost is large, it may become a GRB progenitor since the very low core-envelope coupling has maintained a high angular momentum in the core. 

\section{Evolution of massive stars with mass loss and rotation for 0$< Z \le \sim 0.002$}

\begin{figure}
\includegraphics[width=5.2in,height=2.6in,angle=0]{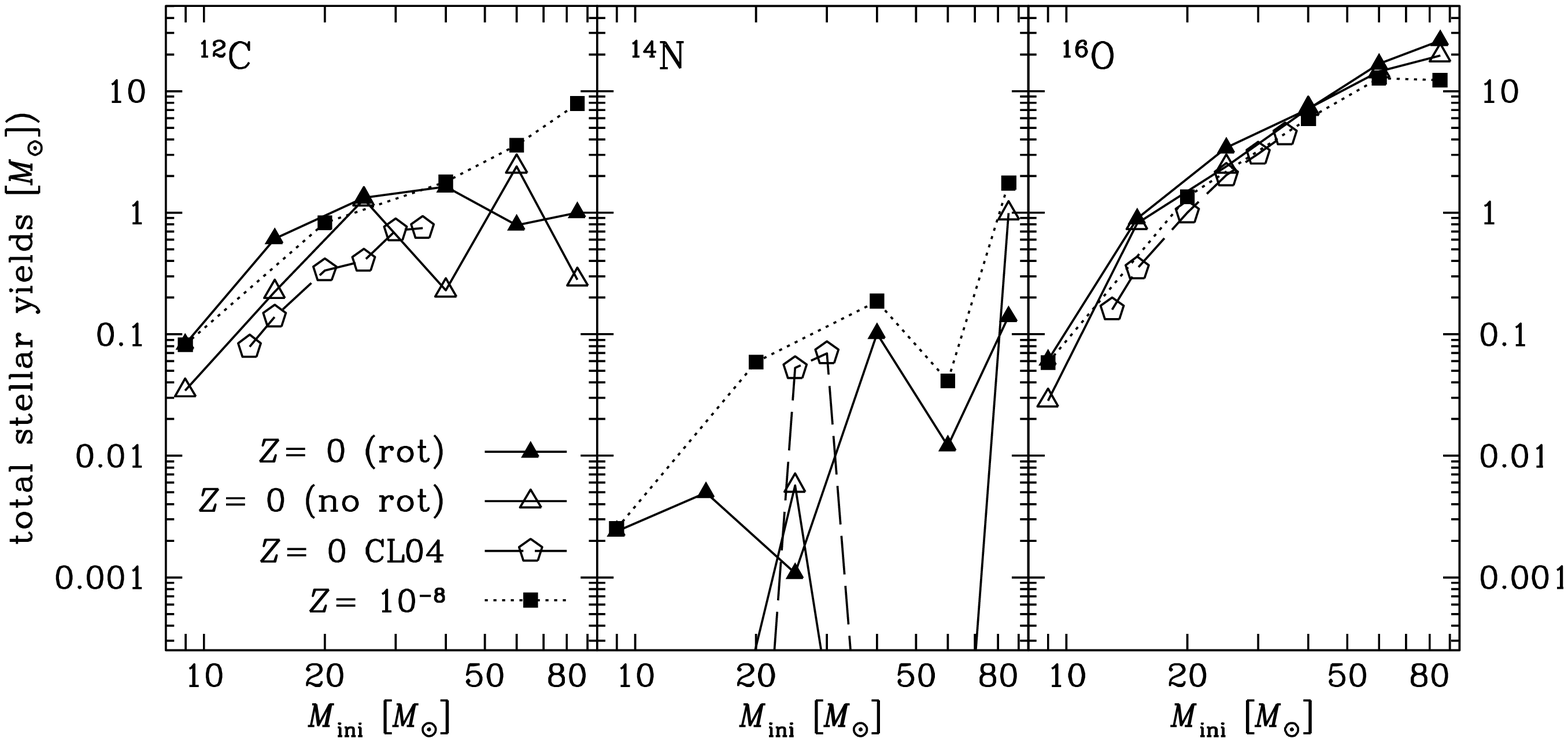}
\caption{Yields comparison between the non-rotating $Z=0$ models from Chieffi \& Limongi (2004; open pentagons), the rotating $Z=10^{-8}$ models from Hirschi (2007; filled squares) and those
of Ekstr\"om et al. (2008b; rotating filled triangles; and non-rotating open triangles) $Z=0$ models. \textit{Left:} $^{12}$C; \textit{centre:} $^{14}$N; \textit{right:} $^{16}$O. Figure taken from Ekstr\"om et al. (2008b).
}\label{ycno}
\end{figure}

\begin{figure}
\includegraphics[width=2.6in,height=2.6in,angle=0]{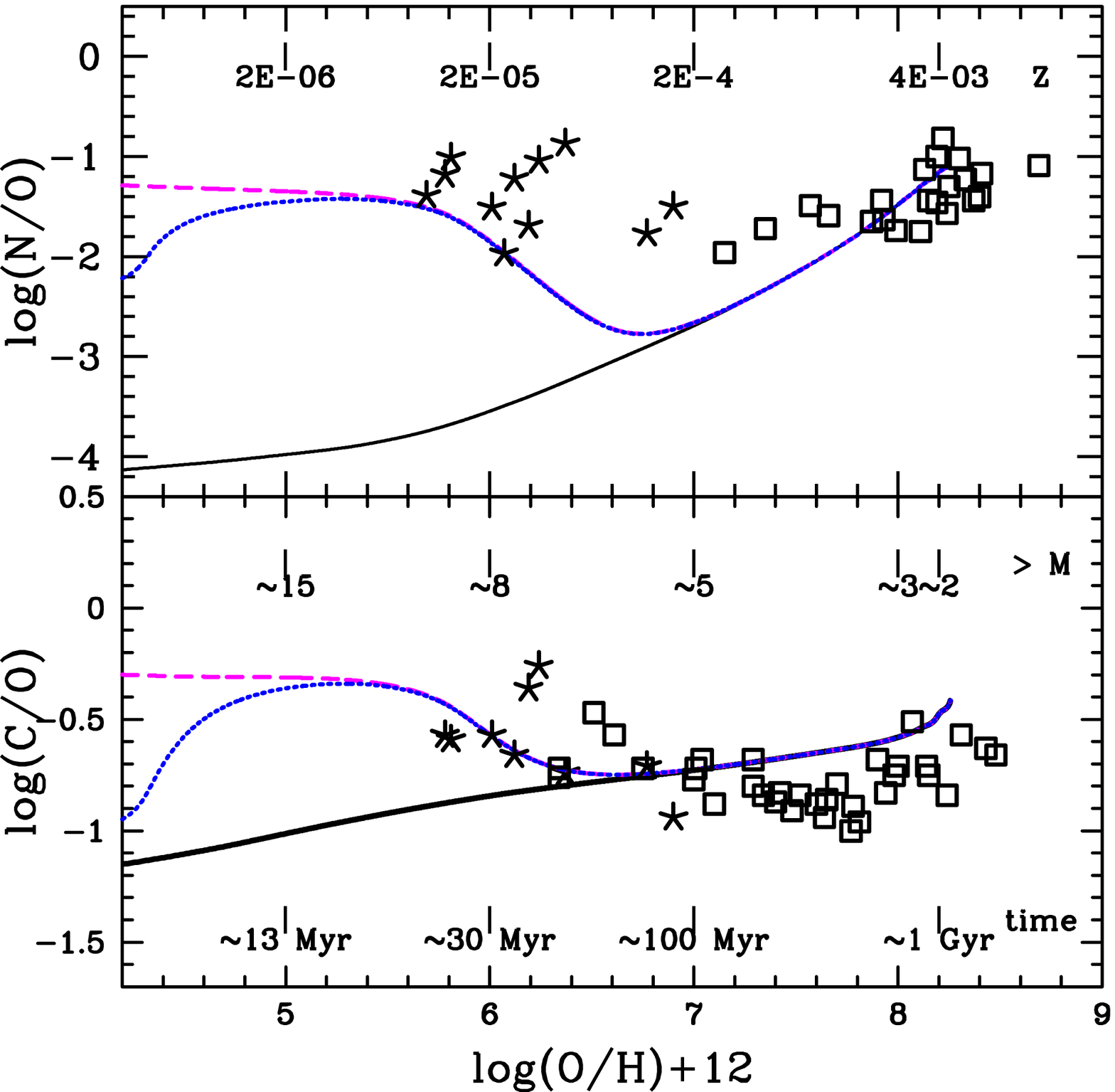}
\hfill
%\hspace{1cm}
\includegraphics[width=2.6in,height=2.6in,angle=0]{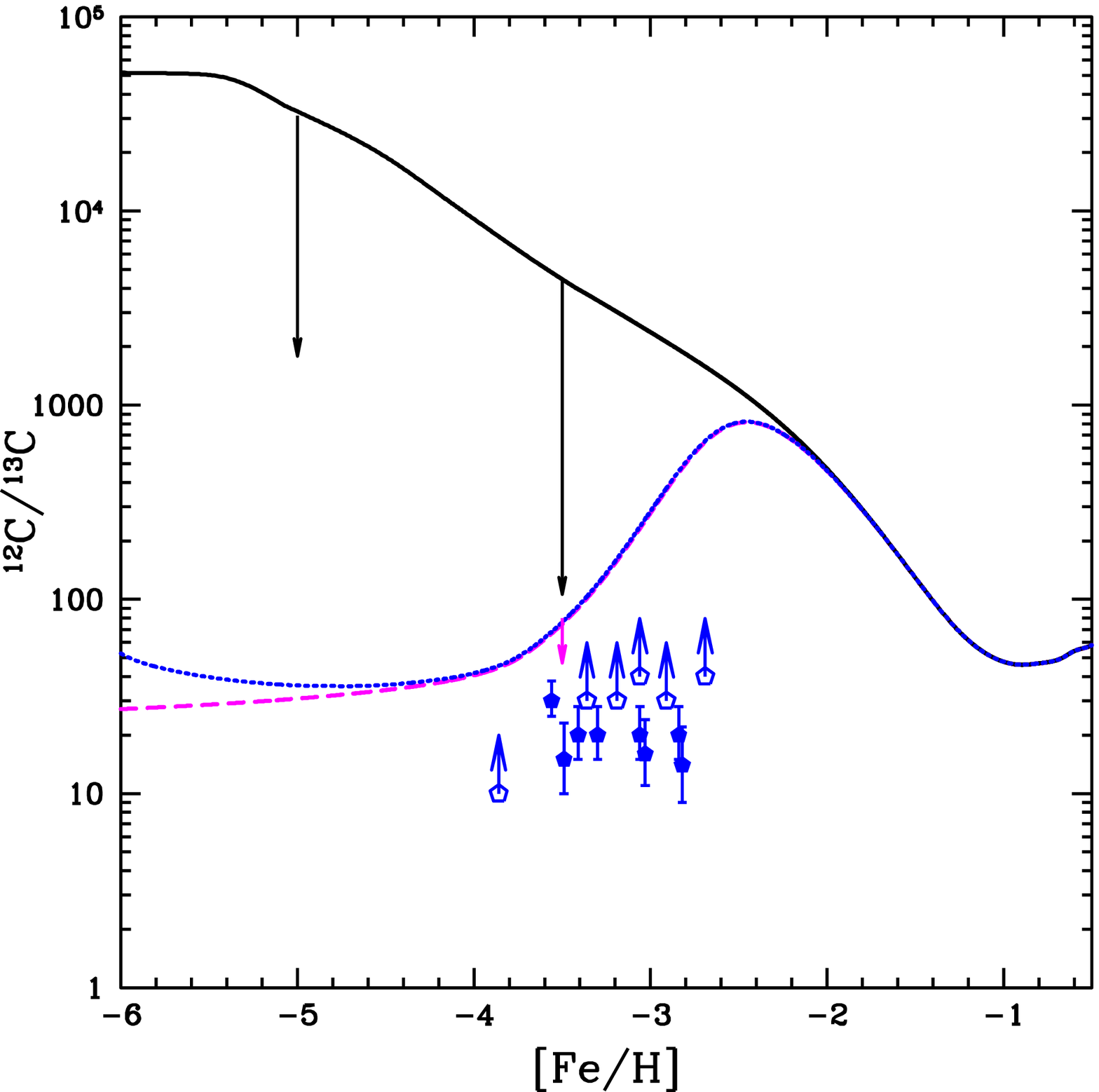}
\caption{The continuous curve is the chemical evolution model obtained with the stellar yields of slow rotating $Z=10^{-5}$ models from Meynet \& Maeder (2002) and Hirschi et al. (2004). The dashed line includes the yields of fast rotating $Z=10^{-8}$ models from Hirschi (2007) at very low metallicity. The dotted curve is obtained using the yields of the $Z=0$ models of Ekstr\"om et al. (2008b) up to $Z=10^{-10}$. \textit{Left:} evolution of the N/O and C/O ratios. Data points are from Israelian et al. (2004; open squares) and Spite et al. (2005; stars). \textit{Right:} evolution of the $^{12}$C/$^{13}$C ratio. Data points are \textit{unmixed} stars from Spite et al. (2006): open pentagons are lower limits. The arrows going down from the theoretical curves indicate the final $^{12}$C/$^{13}$C observed in giants (after the dredge-up), starting from the initial composition values given by the stellar models (see Chiappini et al.
2008). Figure taken from Ekstr\"om et al. (2008b).
}\label{NOC}
\end{figure}

In contrast to the case of massive Pop III stars, which ignite hydrogen through the pp chains,
stars beginning their life with a tiny amount of metals (of the order of 10$^{-10}$ in mass fraction)
ignite their hydrogen through the CNO cycle. The energy produced by the CNO cycle is sufficient to
compensate for the loss of energy by the surface, while that produced by the pp chain is not. 
In this last case, the star has to extract energy from its gravitational reservoir and contracts\footnote{The slow contraction stops when the central temperature becomes high enough for triple alpha reactions to be activated. The $3\alpha$ reactions produce some carbon (of the order of 10$^{-10}$ in mass fraction). From this stage on, hydrogen burns in the core as in more metal rich stars {\it i.e.} through the CNO cycle.}. 
As a consequence the temperature in the core of Pop III stars is in the range of values
for He-burning, while the central temperature in non zero metallicity stars is well below.
This means also that at the end of the MS phase, the core of the Pop III star will not have to
contract a lot in order to activate the reactions of He-burning, while the core of non zero
metallicity stars will have to contract significantly. This contraction will
create a steep gradient
of angular velocity at the border of the core, which will drive a strong mixing, much stronger than
in Pop III stars. 
This is the reason why, 
although the production of primary nitrogen is already quite significant in the rotating Pop III stars, very metal poor stars are still more efficient producers (for a given initial angular momentum content, see Fig.~\ref{ycno}). In Fig.~\ref{NOC} predictions of chemical evolution models using yields of different stellar models are compared with observations of the N/O and
$^{12}$C/$^{13}$C ratios at the surface of metal poor halo stars.

Another consequence of the strong mixing in non-zero metallicity stars is the
evolution toward the red part of the HR diagram (Maeder \& Meynet 2001). In the red part
of the HR diagram, a convective zone appears at the surface which dredges-up at the surface great quantities of CNO elements, leading to the loss of a very large amounts of mass.
In our 60 M$_\odot$ stellar model with $Z=10^{-8}$ and $\upsilon_{\rm ini}=800$ km s$^{-1}$, the CNO
content at the surface amounts to one million times the one the star had at its birth (Meynet et al. 2006). Therefore the global metallicity at the surface becomes equivalent to that of a LMC stars while the star began its life with a metallicity which was about 600 000 times lower! If we apply the same rules used at higher metallicity relating the mass loss rate to the global metallicity, we obtain that the star loses about half
of its initial mass due to this effect. As shown by Meynet et al. (2006) and Hirschi (2007) the matter released
by these winds is enriched in both H- and He-burning products and present striking similarities with the abundance patterns observed at the surface of C-rich Ultra-Metal-Poor-Stars (CRUMPS). This process does not occur in the present Pop III stellar models. As can be deduced from the explanations above, Pop III stars undergo less efficient mixing during the core He-burning phase. Thus the H-burning shell and the outer radiative envelope are enriched at a lower rate in CNO elements. This tends to delay the apparition of an outer convective zone and to reduce its extension. The increase of the surface metallicity remains very modest and does not lead to strong mass loss.

\section{Evolution of massive stars with mass loss and rotation for $Z > \sim 0.002$}

\begin{figure}
\includegraphics[width=2.6in,height=2.6in,angle=0]{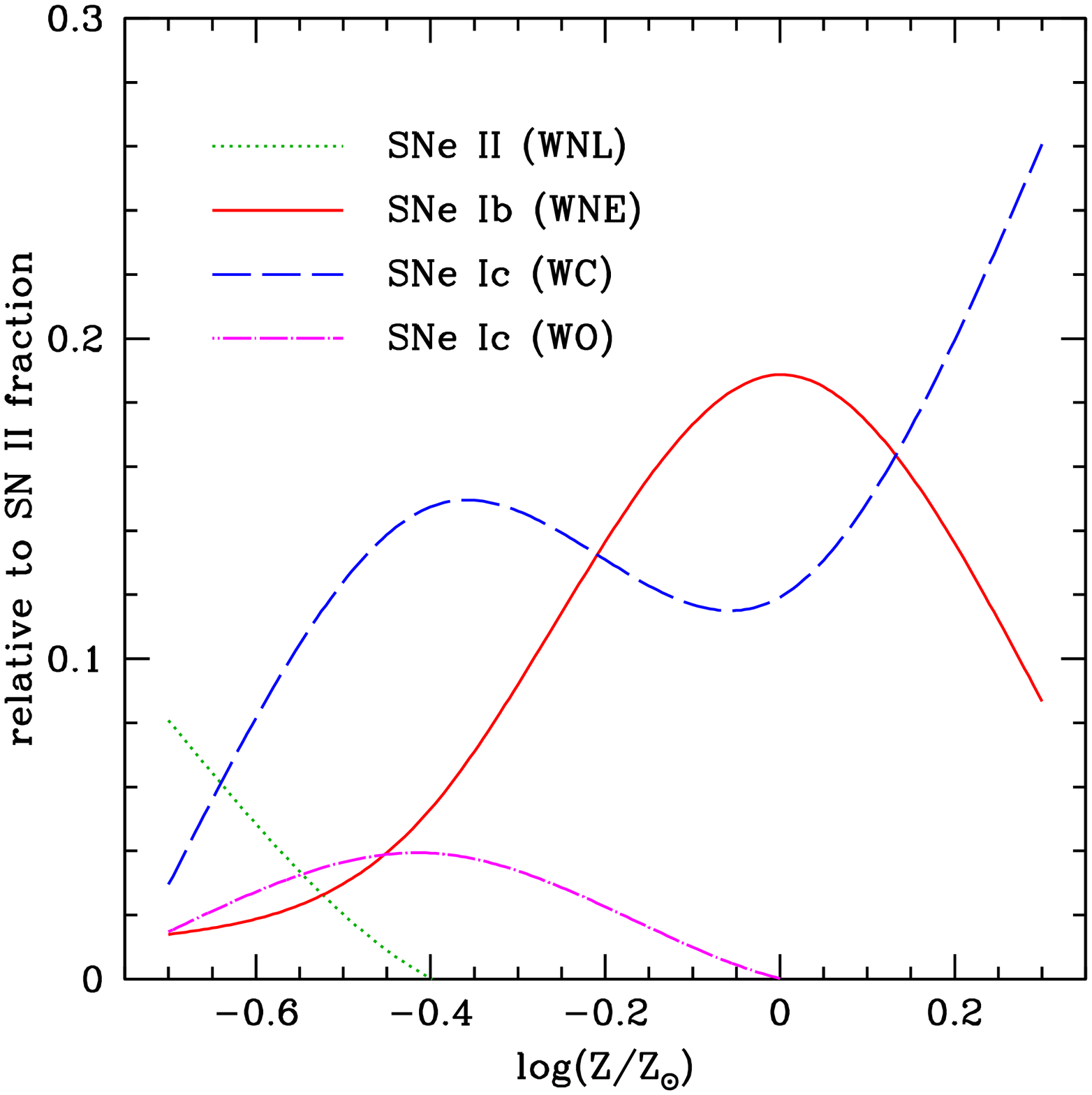}
\hfill
%\hspace{1cm}
\includegraphics[width=2.6in,height=2.6in,angle=0]{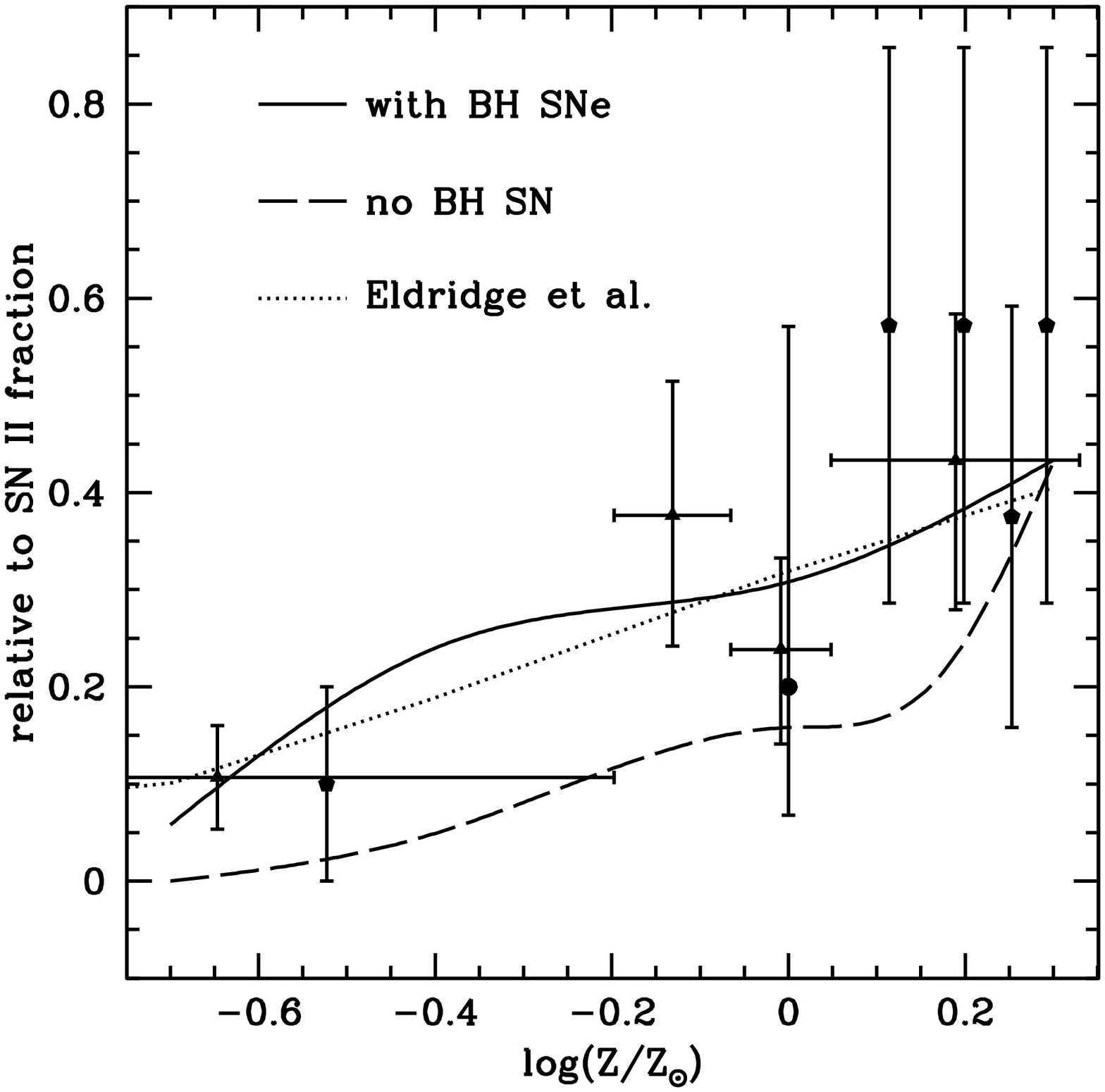}
\caption{{\it Left Panel:} Variation as a function of the metallicity of the number fraction
of supernovae having as progenitors WNL, WNE, WC and WO stars. The fractions were 
deduced from the models of Meynet \& Maeder (2003; 2005) using a Salpeter IMF. {\it Right Panel:}
Rate of SN Ibc / SN II if all models produce a SN (solid line) or if models producing a black holes do not explode in a SN (dashed line). Pentagons are observational data from Prieto et al. (2008), and triangles are data from Prantzos \& Boissier (2003). Figure taken from Georgy et al. in preparation.
}\label{WR}
\end{figure}

Mass loss and rotation play major roles in shaping the populations of massive stars observed in the nearby Universe. For instance predictions of single star models have been obtained for explaining the populations of Be stars (Ekstr\"om et al. 2008a), of blue and red supergiants in the SMC (Maeder \& Meynet 2001), of Wolf-Rayet stars (Meynet \& Maeder 2003, 2005). 
Fig.~\ref{WR} (left panel) shows the variation as a function of the metallicity of the number fraction of supernovae having as progenitors WNL, WNE, WC and WO stars.
We see that WO star progenitors are predicted to occur only at low metallicities. Smith \& Maeder (1991) explained this trend in the following way:
when the mass loss rates are low ({\it i.e.} at low
metallicities), more time is needed to remove the H- and He-rich layers and thus when the He-burning core is uncovered, it has reached a more advanced evolutionary stage ({\it i.e.} more helium converted into carbon and oxygen).
Observations show that indeed most of WO stars (6 out of 8) are found in regions with $Z$ inferior to about 0.9 Z$_\odot$. The fact that WO stars are preferentially found at low metallicity has been invoked to associate these stars to the progenitors of long soft GRBs (Hirschi et al 2005).

Considering that all models ending their lifetime as a WNE or WC/WO phase will explode as a type Ibc supernova, it is possible to compute the variation with the metallicity of the number ratio of type Ibc to type II supernovae. The result is shown in Fig.~\ref{WR} (right panel). One sees that
this ratio increases with the metallicity. This is due to the fact that at higher metallicity, the minimum initial mass of stars ending their life as WNE, or WC/WO stars is lower than at lower
metallicities. Single star model can reasonably well reproduce the observed trend. Models accounting
for single and binary channel (but without rotation) are shown as a dotted line (Eldridge et al. 2008). They also provide a good
fit to the observations. In this last work most of the supernovae originate from the binary channel,
leaving little place for the single star scenario.

It might be that the single and binary channels
will not predict the same distributions of supernovae among type Ib and Ic's. The observations
by Prieto et al. (2008) indicate that type Ic's are about twice as frequent as type Ib's at solar
metallicity. Georgy et al (in preparation) show that rotating single star models can account for this
distribution. It would be very interesting to have the predictions of the binary channel for this feature. Hopefully it will allow to better estimate the respective importance of these two channels.

\end{document}